\documentclass[preprint, superscriptaddress]{revtex4-1}
\usepackage{graphicx}

\usepackage{amstext}
\usepackage{amsmath}
\usepackage{amssymb}
\usepackage{graphicx}
\usepackage{psfrag}
\usepackage{ulem}
\usepackage[inline]{enumitem}
\newcommand{\eq}[1]{Eq.~\eqref{#1}}
\newcommand{\fig}[1]{Fig.~\ref{#1}}

\newcommand{\be}{\begin{equation}}
\newcommand{\ee}{\end{equation}}

\newcommand{\ket}[1]{| #1 \rangle}
\newcommand{\bra}[1]{\langle #1 |}

\begin{document}

\title{Lessons on electronic decoherence in molecules from exact modeling}
\author{Wenxiang Hu}
\affiliation{Materials Science Program, University of Rochester, Rochester, New York 14627, USA}
\author{Bing Gu}
\affiliation{Department of Chemistry, University of Rochester, Rochester, New York 14627, USA}
\author{Ignacio Franco}
\affiliation{Department of Chemistry, University of Rochester, Rochester, New York 14627, USA}
\affiliation{Department of Physics, University of Rochester, Rochester, New York 14627, USA}

\date{\today}

\begin{abstract} 
Electronic decoherence processes in molecules and materials are usually thought and modeled via schemes for the system-bath evolution in which the bath is treated either implicitly or approximately. Here we present computations of the electronic decoherence dynamics of a model many-body molecular system described by the Su-Schreefer-Heeger Hamiltonian with Hubbard electron-electron interactions using an exact method in which both electronic and nuclear degrees of freedom are taken into account explicitly and fully quantum mechanically. To represent the electron-nuclear Hamiltonian in matrix form and propagate the dynamics,  the computations employ a Jordan-Wigner transformation for the fermionic creation/annihilation operators and the discrete variable representation for the nuclear operators. The simulations offer a standard for electronic decoherence that can be used to test approximations. They also provide a useful platform to answer fundamental questions about electronic decoherence that cannot be addressed through approximate or implicit schemes. Specifically, through simulations, we isolate basic mechanisms for electronic coherence loss, and demonstrate that electronic decoherence is possible even for one-dimensional anharmonic nuclear baths. Further, we show that:
\begin{enumerate*}[label=(\roman*)]
\item Decreasing the mass of the bath generally leads to faster electronic decoherence; 
\item Electron-electron interactions strongly affect the electronic decoherence when the electron-nuclear dynamics is not pure-dephasing;
\item Classical bath models with initial conditions sampled from the Wigner distribution accurately capture the short-time electronic decoherence dynamics;
\item Model separable initial superpositions often used to understand decoherence after photoexcitation are only relevant in experiments that employ delta-like laser pulses to initiate the dynamics.
\end{enumerate*} These insights can be employed to interpret coherence phenomenon in molecules and design modeling strategies.
\end{abstract}

\keywords{electronic decoherence | Su-Schreefer-Heeger Hamiltonian | vibronic interactions | exact quantum dynamics}

\maketitle

\section{Introduction}

Electronic decoherence is a basic feature of correlated electron-nuclear states\cite{Izmaylov2017} and accompanies photoexcitation\cite{Franco2008}, passage through conical intersections\cite{Levine2007}, electron transfer\cite{Cheng2009}, or any other dynamical process that creates  superpositions of electronic diabatic states in molecules. Understanding electronic decoherence is central to the description of basic processes such as photosynthesis, vision and electron transport\cite{Levine2007,Cheng2009,Choi2010}, to the development of approximation schemes to the vibronic evolution of molecules,\cite{Kapral2006, Subotnik2011} and to the isolation of superposition states with robust coherence properties that can subsequently be used in quantum technologies\cite{Shapiro2012, Nielsen2011}.

Developing deep insights into electronic decoherence requires a detailed understanding of the dynamics of the nuclear bath that leads to the coherence loss. To see this, consider the  coherence properties of electrons in a general entangled vibronic state 
\begin{equation}
\label{eq:schmidt2}
\ket{\Omega (t)} = \sum_n\ket{E_n} \ket{\chi_n(t)},
\end{equation}
where the $\ket{E_n}$ refer to electronic eigenstates (which are obtained by diagonalizing the Hamiltonian at a particular nuclear geometry) and the $ \ket{\chi_n} $ to the nuclear wavepackets associated with each $\ket{E_n}$. The electronic reduced density matrix associated with such a state is given by
\begin{equation}
\label{eq:rrho}
\hat{{\rho}}_e(t) = \textrm{Tr}_N[\ket{\Omega}\bra{\Omega}]
=\sum_{nm} \langle \chi_m (t) | \chi_n (t) \rangle  | E_n\rangle \langle E_m |,
\end{equation}
where the trace is over the nuclear coordinates. Importantly, note that the coherence between electronic states $n$ and $m$ is governed by the nuclear wavepacket overlap  $S_{nm} =\langle \chi_m(t)|\chi_n(t)\rangle$. Standard measures of decoherence capture precisely this. For instance, the purity
\begin{equation}
\label{eq:purity}
P(t) = \textrm{Tr}[\hat{{\rho}}_e^2(t)] =  \sum_{nm} | S_{nm}|^2 
\end{equation}
decays with the overlaps between the environmental states $S_{nm}$.  Detailed understanding of electronic decoherence can thus be developed by investigating the dynamics of the $S_{nm}$\citep{Izmaylov2017}.

Because of the difficulty in following the vibronic evolution of molecules exactly, most of our insights into electronic decoherence have emerged from models where the nuclear bath is taken into account implicitly or approximately. In implicit approaches, decoherence effects are modeled via master equations\cite{Pachon2012} where the bath is represented using a spectral density with adjustable parameters chosen to reproduce experimental findings. In turn, explicit approaches offer a detailed description of the nuclear dynamics albeit at an approximate level. Approximations that have been employed to model decoherence include surrogate Hamiltonians\cite{Katz2007}, path-integral techniques\cite{Huo2011}, frozen Gaussian approaches\cite{Bedard-Hearn2005}, semi-classical methods\cite{Fiete2003} and quantum-classical methods\cite{Franco2008, Franco2012, Franco2013, Shim2012, ulrich2017, akimov2014, Meier2004}. 

Here we present accurate numerical simulations of electronic decoherence in a model molecular system using an exact method that takes into account both electrons and nuclei explicitly and fully quantum mechanically. As a model, we adopt the Su-Schreefer-Heeger (SSH) Hamiltonian\cite{Heeger1988} for \textit{trans}-polyacetylene  because it captures the essential vibronic phenomena of molecules. To quantify the effects of electron-electron interactions on the electronic decoherence \cite{Kar2016}, we augment this model with a Hubbard electron-electron interaction term\cite{Hubbard238}. This Hamiltonian is represented in matrix form by Jordan Wigner transformation\cite{Jordan1928, simons2010, Flick2016} of the electrons and by discrete variable representation\cite{Colbert1992} (DVR) of the nuclei. The decoherence dynamics of such a system is propagated using the Crank-Nicholson method\cite{Crank1947} for a neutral SSH chain in Fock space with 4 electrons coupled to 2 vibrational modes. The advantage of this method is that it does not invoke any physical approximations, facilitating the interpretation of the results. Further, because it solves the many-body problem exactly, it can access regimes that are challenging for other methods such as those encountered when the nuclear mass is small (where the Born-Oppenheimer approximation and even the very concept of a potential energy surface fails) or when the electron correlation is large. These simulations complement recent efforts to capture  electronic decoherence dynamics in molecules using semiclassical approximations\cite{ulrich2017, Prezhdo1998, Franco2008, Franco2012, Franco2013}, the multiconfiguration time-dependent Hartree method (MCTDH) \cite{Arnold2017, Vacher2017, JanSchulze2016} and a recently proposed generalized theory for the timescale of the electronic decoherence in the condensed phase\cite{Gu2018}.

In addition to providing a standard for electronic decoherence dynamics in closed molecular systems, the exact modeling serves as a platform from which several basic questions about decoherence in molecules can be addressed. We focus this discussion around seven questions that explore the main requirements for the emergence of decoherence, its  basic phenomenology  and the applicability of approximate schemes to capture the decoherence. While the simulations pertain to a particular model system, the generic nature of the employed Hamiltonian permits interpreting the insights that result from this model in a broader sense.

Specifically, we investigate the basic mechanisms for the electronic coherence loss (Sec.~\ref{subsec:mechanisms}) and the bath size requirements for its emergence (Sec.~\ref{subsec:emergency}). We also investigate how timescales of electronic and nuclear decoherences compare (Sec.~\ref{subsec:timescales}), and establish conditions for the accuracy of classical bath models (Sec.~\ref{subsec:accuracy}). In Sec.~\ref{subsec:mass_eff}-\ref{subsec:prep_eff} , we study the effect of changing the nuclear mass and the initial-time preparation method on the electronic decoherence dynamics. Last, in Sec.~\ref{subsec:e_int}, we study the largely unexplained problem of how electronic interactions modulate the electronic decoherence\cite{Kar2016}. We summarize our main findings in Sec.~\ref{sec:Conclusions}.
 	
\section{Model and Methods}
\label{sec:M & M} 

\subsection{Model Hamiltonian}
As an exemplifying model of a molecular system, we adopt the Su-Schrieffer-Heeger (SSH) tight-binding model for \textit{trans}-polyacetylene (PA) augmented with a Hubbard electron-electron interaction term, with Hamiltonian
\begin{equation}
\label{eq:SSH}
\hat{H}_{\text{SSH}} = \hat{H}_{e} + \hat{H}_{N} + \hat{H}_{\text{int}}.
\end{equation}
Here,
\begin{align}
&\hat{H}_{e} = -t_{0}\sum_{n=1}^{N-1}\sum_{\Delta=\pm}(\hat{c}_{n+1,\Delta}^{\dag}\hat{c}_{n,\Delta}+\hat{c}_{n,\Delta}^{\dag}\hat{c}_{n+1,\Delta}) + U\sum_{n=1}^{N}\hat{c}_{n,+}^{\dag}\hat{c}_{n,+}\hat{c}_{n,-}^{\dag}\hat{c}_{n,-} \nonumber, \\
&\hat{H}_{N} = \sum_{n=1}^{N}\frac{\hat{{p}}_{n}^2}{2M}+\frac{K}{2}\sum_{n=1}^{N-1}(\hat{{u}}_{n+1}-\hat{{u}}_{n})^2\nonumber, \\ 
&\hat{H}_{\text{int}} = \alpha\sum_{n=1}^{N-1}\sum_{\Delta=\pm}(\hat{c}_{n+1,\Delta}^{\dag}\hat{c}_{n,\Delta}+\hat{c}_{n,\Delta}^{\dag}\hat{c}_{n+1,\Delta})(\hat{u}_{n+1}-\hat{u}_{n}) \nonumber,
\end{align}
are the electronic, nuclear and electron-nuclear interaction components of the Hamiltonian, respectively. In this model, each site $n$ represents a CH unit along a PA chain with $N$ repeating units. In the electronic component, the fermion creation (annihilation) operators $\hat{c}_{n,\Delta}^{\dag}(\hat{c}_{n,\Delta})$ creates (or annihilates) an electron on site $n$ with spin $\Delta$, and are subject to the usual fermionic anti-commutation rules. The electronic Hamiltonian describes hopping process of the $\pi$ electrons with same spin between different sites with hopping strength $t_{0}$, and a Hubbard electron-electron repulsion ($U \geqslant 0$) when two electrons occupy the same site. In the nuclear component, the $\hat{u}_{n}$ refers to the displacement of the $n$th CH unit from position $na$ where $a$ is the lattice constant, $\hat{p}_{n}$ is the momentum conjugate to $\hat{u}_{n}$, $M$ is the mass of the CH group and $K$ is the effective spring constant. The electron-nuclear interaction modulates the hopping integral as neighboring nuclei come closer together.  In this paper, we use the standard SSH parameters for PA: $\alpha=4.1\,\text{eV}/\text{\AA}$, $K=21\,\text{eV}/\text{\AA}^{2}$, $t_{0}=2.5\,\text{eV}$, $M=1349.14\,\text{eV}\text{fs}^2/\text{\AA}^2$ and $a=1.22\,\text{\AA}$. The quantity $U$ is taken to be $U=0$ except in Sec.~\ref{subsec:e_int} where the effect of the electronic correlations on the decoherence is investigated.

Below, we investigate the exact vibronic dynamics for a neutral SSH chain with 4 electrons. The end atoms of the molecule are taken to be clamped, resulting in two vibrational modes;  a high frequency C-C stretching optical mode where the two middle nuclei move in opposite direction with identical amplitude, and a lower frequency acoustic mode where two middle nuclei move in the same direction with identical amplitude.  Because the vibronic coupling is proportional to $(\hat{u}_n - \hat{u}_{n-1})$, the optical mode couples strongly to the electronic degrees of freedom while the acoustic mode is weakly coupled.  

\subsection{Matrix representation of the Hamiltonian}
The matrix representation of the SSH Hamiltonian is constructed by tensor product of its electronic and nuclear components. To represent the fermionic creation and annihilation operators in matrix form, we adopt the Jordan-Wigner Transformation\cite{Jordan1928, simons2010, Flick2016}. In this transformation, the fermionic annihilation operators at site $n$ with spin $\Delta$ can be represented by a string matrix $e^{i\phi_{n, \Delta}}$ times the corresponding spin-$\frac{1}{2}$ Pauli lowering matrix $\sigma_{n, \Delta}$ at the same site and with the same spin, \textit{i.e.} $\hat{c}_{n,\Delta} \doteq e^{i\phi_{n, \Delta}}\sigma_{n, \Delta}$. Similarly, the fermionic creation operator is represented as $\hat{c}_{n,\Delta}^{\dag} \doteq \sigma_{n, \Delta}^{\dag}e^{-i\phi_{n, \Delta}}$, where $\sigma_{n, \Delta}^{\dag}$ is the spin-$\frac{1}{2}$ Pauli raising matrix at site $n$ with spin $\Delta$. Here, the phase matrix $\phi_{n, \Delta}$ contains the sum over all the occupation matrices to the left of ($n, \Delta$), \textit{i.e.} $\phi_{n, \Delta} = \pi \sum_{(k, \Delta') < (n, \Delta)}\sigma_{k, \Delta'}^{\dag}\sigma_{k, \Delta'}$. In our notation, we assume that the spin up $(+)$ is at the left of spin down $(-)$ for each site.

The spin number operator $\sigma_{n, \Delta}^{\dag}\sigma_{n, \Delta}$ is idempotent. Further, $\sigma_{n, \Delta}^{\dag}\sigma_{n, \Delta}$ with different ($n, \Delta$) commute. Thus, by Taylor expanding each component of $e^{i\phi_{n, \Delta}}=\prod_{(k, \Delta') < (n, \Delta)}e^{i\pi\sigma_{k, \Delta'}^{\dag}\sigma_{k, \Delta'}}$, the fermionic annihilation operators can be expressed as:
\begin{widetext}
\be
\label{eq:giant_matrix}
\begin{array}{ccccccccccccc} 
\hat{c}_{1+} & \doteq & \sigma_{1+} & \bigotimes & I_{1-} & \bigotimes& {I}_{2+} &\bigotimes & \cdots & \bigotimes & {I}_{N+} & \bigotimes & {I}_{N-} \\
\hat{c}_{1-} & \doteq & 1-2\sigma_{1+}^{\dag}\sigma_{1+} & \bigotimes & \sigma_{1-} & \bigotimes& {I}_{2+} &\bigotimes & \cdots & \bigotimes & {I}_{N+} & \bigotimes & {I}_{N-} \\
\hat{c}_{2+} & \doteq & 1-2\sigma_{1+}^{\dag}\sigma_{1+} & \bigotimes & 1-2\sigma_{1-}^{\dag}\sigma_{1-} & \bigotimes& \sigma_{2+} &\bigotimes & \cdots & \bigotimes & {I}_{N+} & \bigotimes & {I}_{N-} \\
\vdots\\
\hat{c}_{N-} & \doteq & 1-2\sigma_{1+}^{\dag}\sigma_{1+} & \bigotimes & 1-2\sigma_{1-}^{\dag}\sigma_{1-} & \bigotimes& 1-2\sigma_{2+}^{\dag}\sigma_{2+} &\bigotimes & \cdots & \bigotimes & 1-2\sigma_{N+}^{\dag}\sigma_{N+} & \bigotimes & \sigma_{N-} 
\end{array},
\ee
\end{widetext}
where $I_i$ is the $2\times2$ identity matrix in the $i$th subspace. The fermionic creation operators at site $n$ with spin $\Delta$ can be represented simply by replacing $\sigma_{n, \Delta}$ by $\sigma_{n, \Delta}^{\dag}$  in \eq{eq:giant_matrix}. In this way, the resulting operators satisfy the desired fermionic anti-commutation rules ($\{\hat{c}_{n,\Delta}, \hat{c}_{n',\Delta'}^{\dag}\} = \delta_{nn'}\delta_{\Delta\Delta'}, \{\hat{c}_{n,\Delta}, \hat{c}_{n',\Delta'}\} = \{\hat{c}_{n,\Delta}^{\dag}, \hat{c}_{n',\Delta'}^{\dag}\} = 0$) because spin-$\frac{1}{2}$ Pauli matrices satisfy the following relations: $\{\sigma_{n,\Delta}, \sigma_{n,\Delta}^{\dag}\} = 1$, $\{1-2\sigma_{n,\Delta}^{\dag}\sigma_{n,\Delta}, \sigma_{n,\Delta}^{\dag}\} = 0$ and $\{1-2\sigma_{n,\Delta}^{\dag}\sigma_{n,\Delta}, \sigma_{n,\Delta}\} = 0$. 

In this way, we can represent the total electronic Fock space by tensor products of $2\times2$ Hilbert spaces for each site and spin $\mathcal{H}_{i}^{2\times2}$ as: 
\begin{widetext}
\be
\label{eq:Hilbert_tesor_product}
\begin{array}{ccccccccccccc}
\mathcal{H}_{1+}^{2\times2} & \bigotimes & \mathcal{H}_{1-}^{2\times2} & \bigotimes & \mathcal{H}_{2+}^{2\times2} & \bigotimes & \cdots & \bigotimes & \mathcal{H}_{N+}^{2\times2} & \bigotimes &  \mathcal{H}_{N-}^{2\times2} 
\end{array}.
\ee
\end{widetext}
This Fock space includes all possible electronic number states. To reduce the computational effort, we project the electronic Fock space to a Hilbert space with a fixed $n_{e}$ number of electrons. This is possible since the SSH Hamiltonian commutes with the electron number operator $\hat{N}_{e} = \sum_{n=1}^{N}\sum_{\Delta=\pm}\hat{c}_{n,\Delta}^{\dag}\hat{c}_{n,\Delta}$,
\be
[\hat{H}_{\text{SSH}},\hat{N}_{e}]=0,
\ee 
and thus the dynamics preserves $n_{e}$. The size of the net electronic basis is $2^{2n_{e}}$ and can represent any many-body state in the electronic system with a fixed number of electrons $n_{e}$.

To represent the nuclear operators in terms of matrices, we employ the Discrete Variable Representative (DVR) as proposed in Ref.~\onlinecite{Colbert1992}. For simplicity, we elucidate this method with one nuclear degree of freedom. For this degree of freedom, a basis consisting of grid points, $\{\ket{i}\}$, is employed.  The matrix elements of the kinetic energy operator in this basis are:
\be
\label{K_DVR}
\bra{i}\hat{T}\ket{i'} = \frac{\hbar^2(-1)^{i-i'}}{2M(\Delta x)^2}\left\{
       \begin{array}{cccccc}
       \pi^2/3&, & \quad   &i&=&i' \\
       \frac{2}{(i-i')^2}&,& \quad &i&\neq& i'
       \end{array}
\right\},
\ee
where $\Delta x$ is the grid spacing. Correspondingly, the matrix elements of the position dependent function $V(\hat{u})$ are:
\be
\label{V_DVR}
\bra{i}V(\hat{u})\ket{i'} = V(u(i))\delta_{ii'}.
\ee
This idea can be extended to many degrees of freedom and is used to represent the nuclear component of the SSH Hamiltonian in matrix form.

DVR methods have been proved to be highly accurate to solve a variety of problems in molecular quantum dynamics and vibration-rotation spectroscopy\citep{Light2007}. In this study with two nuclear degrees of freedom, the method provides convergence with respect to the grid spacing at $\Delta x = 0.02 \text{\AA}$ for both degrees of freedom. These results cannot be achieved by the general grid basis method in which the second derivative in kinetic term is represented by a tridiagonal matrix because this method require much smaller grid spacing leading to large memory needs. All results presented here have been tested for convergence in the grid spacing and the range of space considered in the simulation.

\subsection{Dynamical propagation}
To propagate the dynamics we employ Crank-Nicholson scheme\cite{Crank1947}. In it, the time operator $\hat{U}(t+\Delta t, t)$ is computed by a first order $\mathrm{Pad\acute{e}}$ approximation\cite{Baker} as 
\be
\label{pade}
\hat{U}(t+\Delta t, t) = \frac{1-i\frac{\Delta t}{2}\hat{H}(t+\frac{\Delta t}{2})}{1+i\frac{\Delta t}{2}\hat{H}(t+\frac{\Delta t}{2})}.
\ee 
In this way, the propagation from $\ket{\Psi(t)}$ to $\ket{\Psi(t+\Delta t)}$ is transformed into the solution of the linear equation
\begin{equation}
\label{eq:ck_linear}
\hat{L}\ket{\Psi(t+\Delta t)}=\ket{b},
\end{equation}
where 
\begin{align}
\label{eq:ck_Lb}
& \hat{L} = 1+i\frac{\Delta t}{2}\hat{H}(t+\frac{\Delta t}{2}), \\
& \ket{b} = (1-i\frac{\Delta t}{2}\hat{H}(t+\frac{\Delta t}{2}))\ket{\Psi(t)}.
\end{align}
This linear equation is solved using a biconjugate gradient stabilized iterative method\cite{Vorst1992} which is stable and faster than direct methods such as Gaussian elimination. Unless specified otherwise, results will be reported with $\Delta t=0.01$ fs. For the model with standard SSH parameters prepared in separable superpositions states, this time step offers converged results (within $2\%$ error) up to $\sim500~\text{fs}$ as verified by performing the dynamics with $\Delta t=0.001$ fs. For larger times, it offers qualitatively correct dynamics. 

\section{Results and Discussion}
\label{sec:R & D}
The decoherence dynamics is investigated by following the purity $P(t)$ [\eq{eq:purity}]. The quantity $P=1$ for a pure state, $P<1$ for a mixed state and $P=1/M$ for a maximally entangled state of $M$ levels with equal populations. Since the vibronic dynamics is solved exactly, the simulations take into account all possible potential energy surfaces (PESs) for the molecule and can access regimes where the Born-Oppenheimer picture or mixed quantum-classical schemes are inadequate. 

\begin{figure}[htbp]
\includegraphics[width=0.5\textwidth]{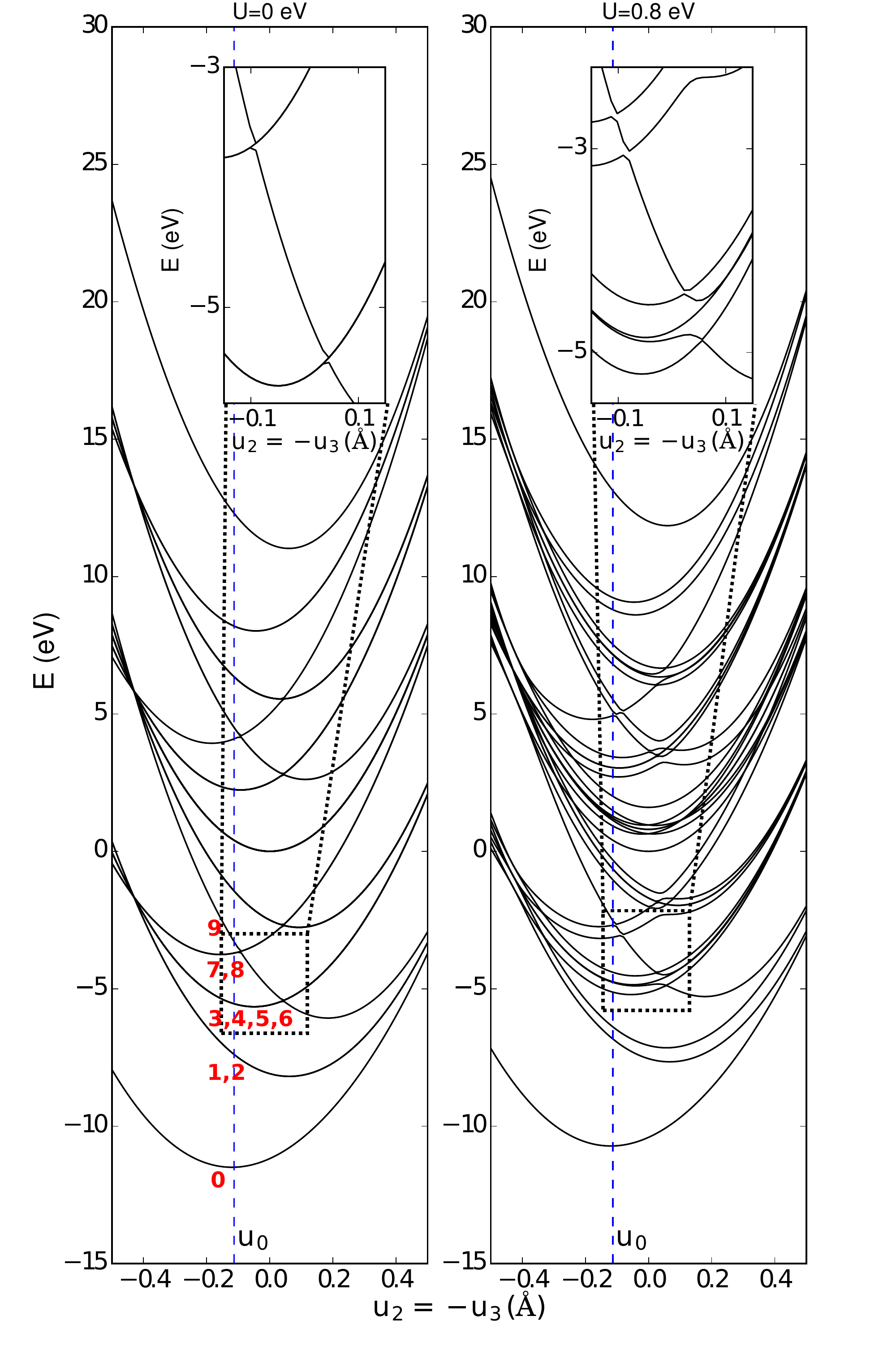}
\caption{Slice of the potential energy surfaces of the SSH Hamiltonian along $u_2= - u_3$ with electron-electron interactions $U=0$ eV (left) and $U=0.8$ eV (right). Here, $u_0$ is the ground-state equilibrium geometry. The numbers 0-9 (in red) are used to label  the first 10 electronic states. For definitiveness, the figure only shows those states with zero net spin along the $z$ direction.}
\label{fig:pes}
\end{figure}

The following analysis is framed around seven questions on the electronic decoherence that are addressed in the light of the exact method.

\subsection{What are the main mechanisms for the electronic coherence loss?}
\label{subsec:mechanisms}

There has been significant discussion in the literature about the main mechanisms for electronic decoherence in molecules\cite{Prezhdo1998, Gu2017, Franco2012, Franco2013, Arnold2017, Vacher2017, Fiete2003, Kar2016, Izmaylov2017, Pachon2012}. From \eq{eq:purity}, it is clear that the decoherence arises due to the nuclear wavepacket evolution in alternative diabatic PESs that lead to a decay in the nuclear wavepacket overlaps. A recently proposed theory of the electronic decoherence shows that, in the short time, this can be divided into pure-dephasing dynamics, transitions between diabatic states and their interference\cite{Gu2018}. 

To connect with these previous efforts and illustrate the main mechanisms for coherence loss in this model, we first consider the case in which the SSH chain is prepared in a separable tensor product of the form
\be
\label{eq:init0}
\ket{\Omega(t=0)} = \frac{1}{\sqrt{2}}(\ket{E_0(u_{0})} + \ket{E_1(u_{0})})\otimes \ket{\chi(t=0)},
\ee
where the electrons are initially in a superposition state  of the ground and the first excited electronic state determined at the equilibrium nuclear coordinates $u_{0}$ of the ground PES, while the initial nuclear state $\ket{\chi(t=0)}$ is taken to be the ground vibrational state of the ground PES. As shown in \fig{fig:pes} (left panel), these two PESs are well separated in energy and there are no avoided crossings, making the pure-dephasing model of the decoherence applicable. In this model, the vibronic evolution leads to an entangled vibronic state of the form
\be
\label{eq:super1}
\ket{\Omega(t)} = \frac{1}{\sqrt{2}}(\ket{E_0(u_{0})}\otimes \ket{\chi_0(t)} + \ket{E_1(u_{0})}\otimes \ket{\chi_1(t)}),
\ee
with initial condition $\ket{\chi_0(t=0)} = \ket{\chi_1(t=0)} = \ket{\chi(t=0)}$. For this particular case, since $|\chi_0(t)\rangle$ is stationary, $P = \frac{1}{2} + \frac{1}{2}|\langle \chi_0(t)|\chi_1(t)\rangle |^2 = \frac{1}{2} + \frac{1}{2}|\langle \chi_0(0)|\chi_1(t)\rangle |^2 = \frac{1}{2} + \frac{1}{2}|\langle \chi_1(0)|\chi_1(t)\rangle |^2 = \frac{1}{2} + \frac{1}{2}|A(t)|^2$, where $A(t) = \langle \chi_1(0)|\chi_1(t)\rangle $ is the autocorrelation function of the excited state nuclear wavepacket. Thus, $P(t)$ is determined by $A(t)$ provided that the dynamics is pure-dephasing. The autocorrelation function can be computed without propagating the quantum state as:
\begin{align}
\label{eq:autocorrelation}
|A(t)|^2 &= |\langle \chi_1(t=0)|\chi_1(t)\rangle |^2 \\
&= \sum_{n,m=0}^{\infty} |a_n|^2 |a_m|^2 e^{i(\epsilon_n-\epsilon_m)t/\hbar}\nonumber,
\end{align}
where $\{\epsilon_n\}$ is the vibrational energy spectrum of the first excited PES and $\{a_n=\langle \phi_n|\chi_1(0)\rangle\}$ are the components of the initial state $|\chi_1(t=0)\rangle$ projected along the vibrational eigenstates $\{|\phi_n\rangle\}$ of the excited PES.

\begin{figure}[htbp]
\includegraphics[width=0.8\textwidth]{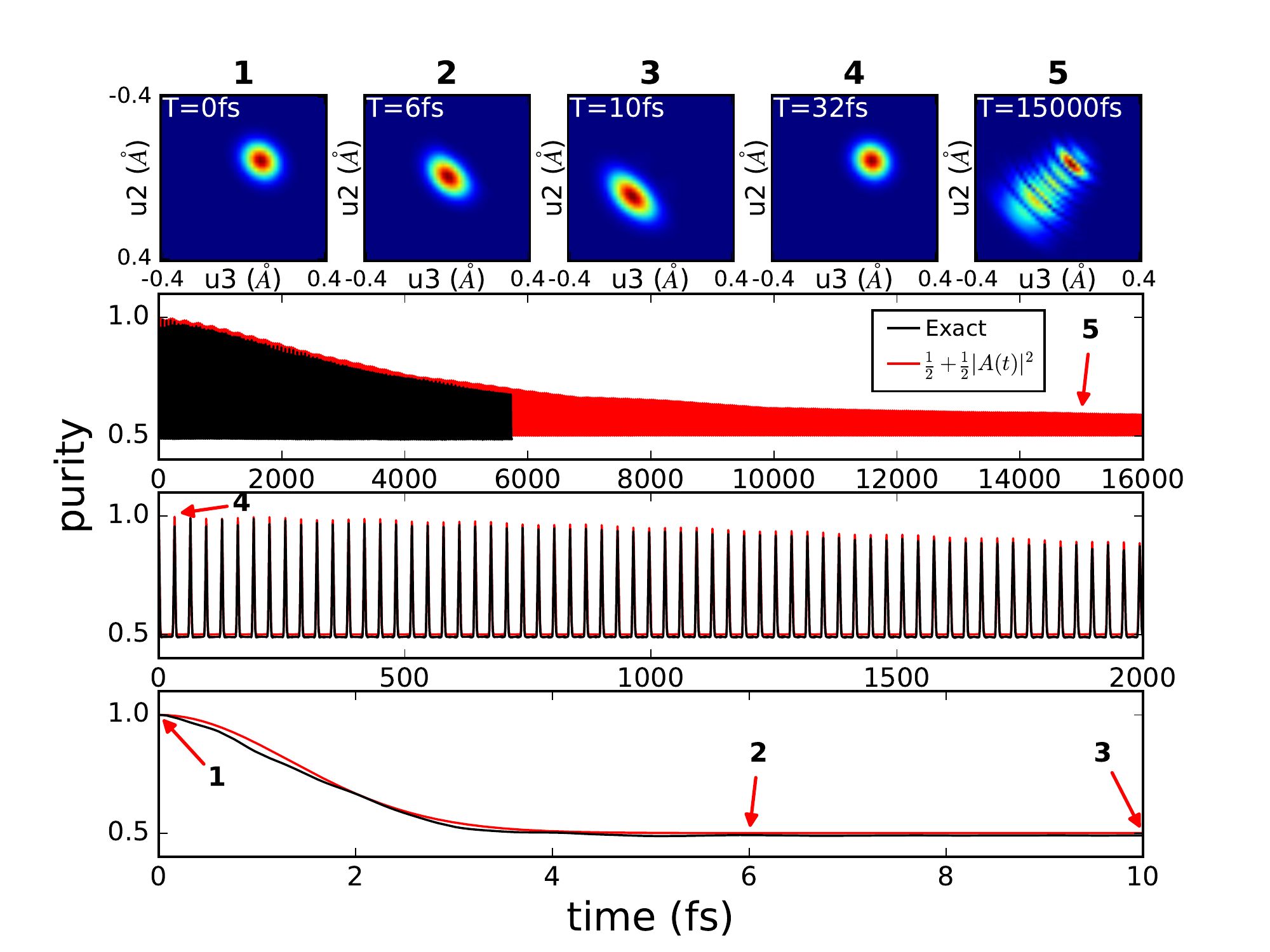}
\caption{Decoherence dynamics for a neutral SSH chain with 4 electrons. The plots show the purity dynamics for a chain initially prepared in the superposition state with equal coefficients between the ground and first excited state in \eq{eq:init0} using a time-step $\Delta t = 0.001~\text{fs}$. The decay and recurrences in the purity reflect the vibrational dynamics in the excited anharmonic state PES. Snapshots of the nuclear probability density in the excited diabatic state are shown in the upper panels. The purity dynamic assuming a pure-dephasing model is shown in red.}
\label{fig:exact}
\end{figure}

As shown in \fig{fig:exact}, the purity decay computed by the autocorrelation function is in quantitative agreement with the decoherence dynamic obtained via dynamics propagation with time step $\Delta t = 0.001\, \text{fs}$ of the chain starting from the superposition state in \eq{eq:init0}. For this time step the dynamics is essentially exact. Thus, a pure-dephasing picture is valid to illustrate the main mechanisms of the electronic coherence loss when starting from the state in \eq{eq:init0}.

\begin{figure}[htbp]
\includegraphics[width=0.5\textwidth]{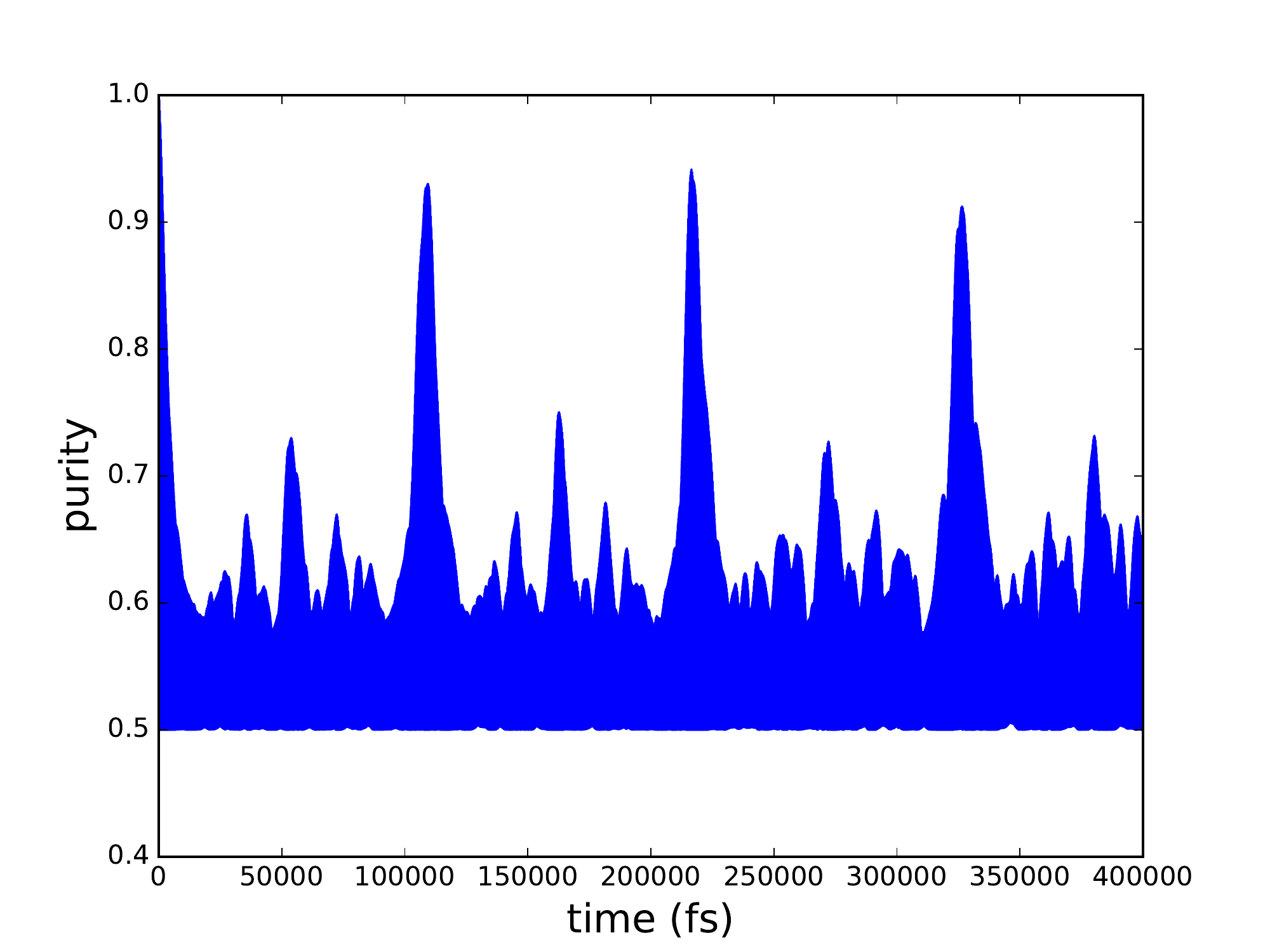}
\caption{Fractional revivals in the decoherence dynamics of a SSH chain for the initial state in \fig{fig:exact}. The decoherence was computed through the autocorrelation function as $P = \frac{1}{2} + \frac{1}{2}|A(t)|^2$.}
\label{fig:revivals}
\end{figure}

We identify three distinct regions for the electronic decoherence. In the first $4~\text{fs}$, the purity exhibits a Gaussian decay that arises due to the initial wavepacket motion on the excited state diabatic PES. Such initial dynamics leads to a decay in the nuclear overlap $|\langle \chi_0(0)|\chi_1(t)\rangle |^2$ as the snapshots of the nuclear wavepacket at $t=6$ fs and $ 10$ fs show. This decay of wavepacket overlap is the mechanism for the short-time decoherence that is dominant in condensed phase environments where recurrences are not expected, and that is captured by theories for decoherence timescales\cite{Prezhdo1998, Gu2017, Gu2018}. Nevertheless, in this model, since the excited PES is bounded and of low dimensionality, the nuclear wavepacket eventually returns to its starting point leading to a recurrence in the purity corresponding to the snapshot at $t=32$ fs. However, due to the anharmonicity of the first excited PES, the recurrences in the purity are never complete and this leads to an overall decay in the purity of the system. For longer times, the nuclear wavepacket is spread along the optical mode of the PES as the snapshot at $t=15000$ fs illustrates, which leads to a  decay in purity with small high frequency oscillations. Thus, this model recovers the initial Gaussian decay of purity that dominates the electronic decoherence in the condensed phase\cite{Arnold2017, Vacher2017, Prezhdo1998, Gu2017}. Further, it captures recurrences that can be observed in small molecular systems and their decay in this case due to anharmonicities in the PES. At even longer times, beyond these two regions, fractional revivals of purity are observed (see \fig{fig:revivals}) which is consistent with quantum revival theory\cite{Robinett2004}. This long time behavior is beyond the applicability of the dynamics method used here, but can be estimated using \eq{eq:autocorrelation} as was done in \fig{fig:revivals}. In this paper, we focus on the first two regions since they are expected to be the most relevant for molecules\cite{Arnold2017, Vacher2017}. 

Interestingly, the recurrence structure discussed above has also been encountered in experiments investigating the vibrational wavepacket evolution of $\text{Br}_2$ molecules in the presence of a solid $\text{Ar}$ environment via ultrafast pump-probe spectroscopy\cite{Guhr2004} and in those studying the evolution of Rydberg electronic wavepackets in $\text{K}$ atom by photoionization measurements\cite{stroud1990}.

\subsection{How large should a bath be in order for decoherence to emerge?} 
\label{subsec:emergency}
Another basic question in electronic decoherence is to determine the size of the bath required for decoherence to emerge. Investigations for a spin coupled to a bath of spins\cite{Schlosshauer2005} indicate that a bath as small as 20 spins is sufficient to generate a Gaussian decay in the overlap of the environmental states.  The question is: in a typical molecule, is decoherence only salient in the condensed phase where the electrons couple to a macroscopic number of bath degrees of freedom or are a few vibrational coordinates enough, and if so, how many? 

As shown in \fig{fig:exact}, two vibrational degrees of freedom are enough to generate decoherence. However, as discussed below, in fact, just one vibrational coordinate is enough for decoherence to emerge since only the optical mode plays an important role during the dynamics.  

\begin{figure}[htbp]
\centering
\includegraphics[width=0.5\textwidth]{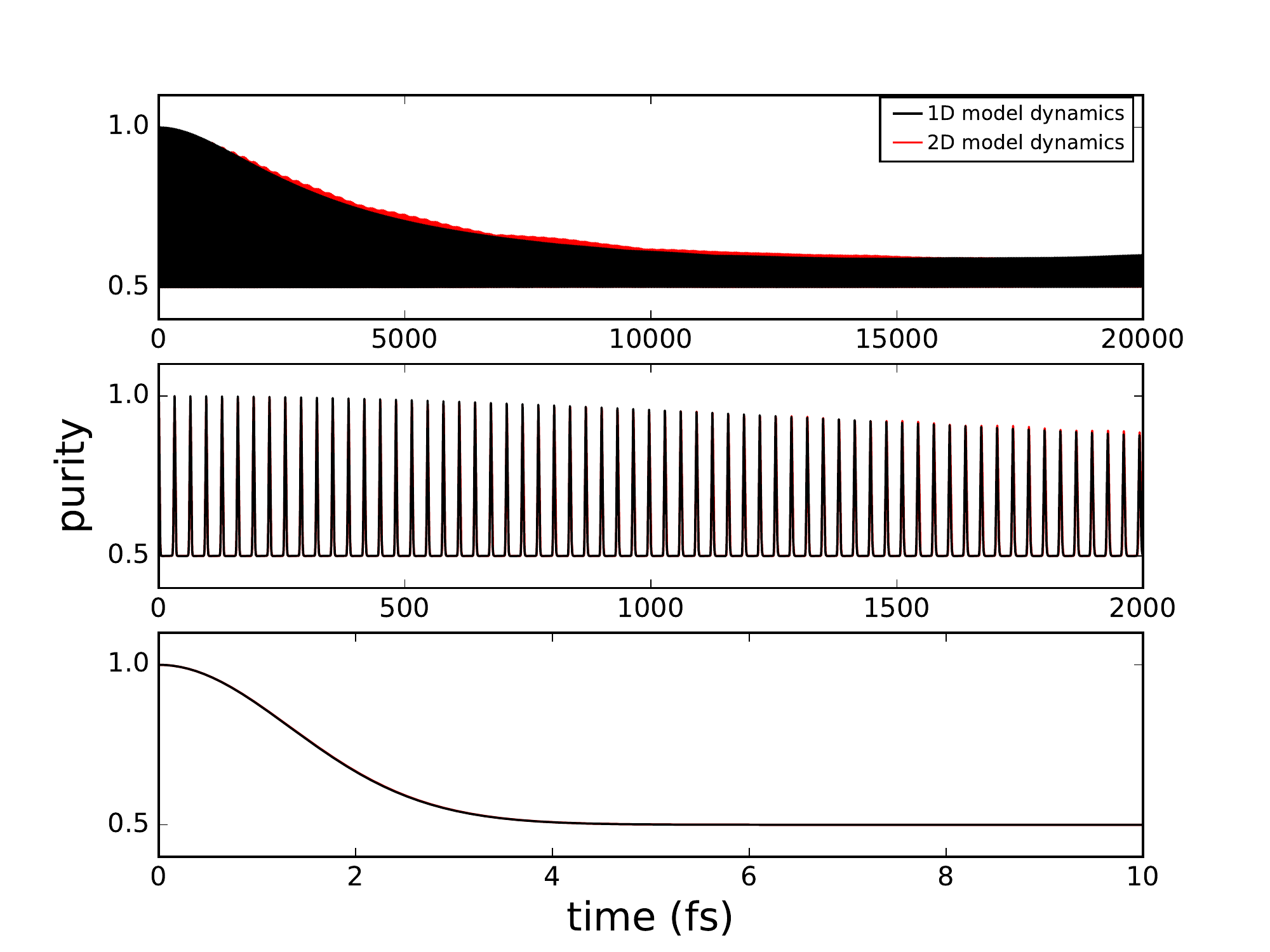}
\caption{The purity decay along the $u_2=-u_3$ direction (optical mode) on the first excited PES (1D model, black) and along the full first excited PES (2D model, red). The red line cannot be seen as it coincides exactly with the 1D model black line. This implies that the nuclear evolution along the optical mode dominates the decoherence dynamics.}
\label{fig:fitting}
\end{figure} 

To see this, consider \fig{fig:fitting} which shows the purity decay computed through the autocorrelation function along the optical mode on the first excited PES (1D model) and that on the full first excited PES (2D model). In this figure, the purity of the two models essentially coincides. Thus, we conclude that the coupling of the electrons to one vibrational degree of freedom (in this case the optical high frequency mode) is sufficient to induce the electronic decoherence. While a macroscopic condensed phase is not required for the emergence of electronic coherence loss, a larger number of bath degrees of freedom will prevent the emergence of partial recurrences in the purity, as observed in Ref.~\onlinecite{Arnold2017, Vacher2017}.

\subsection{How do the timescales for electronic and vibrational decoherence compare?}
\label{subsec:timescales}
Conventional wisdom indicates that electronic decoherence is fast ($\sim 10$ fs) while vibrational decoherence is slow ($\sim 10^2-10^3$ fs)\cite{Fleming1990}. However, in this model the electronic decoherence rate  is identical to the nuclear one, i.e. $P_{e}(t) = P_{N}(t)$ . This is a consequence of the Schmidt theorem\cite{Schmidt1907, Nielsen2011} (or the Carlson-Keller theorem\cite{Carlson1961}) which indicates that for closed system-bath systems the purity of the system and the bath coincide (see Ref.~\onlinecite{Izmaylov2017} for a simple derivation of this fact). What is the origin of this apparent discrepancy?

To resolve this, consider a molecule in the presence of solvent. For short times, the total purity decay of molecular vibrational degrees of freedom is just the product of purity decay due to entanglement with electrons and with solvent\cite{Gu2017}:
\be
P_N(t) = P_{N-e}(t)P_{N-s}(t) = \text{exp}(-\frac{t^2}{\tau_{d}^{(e)2}})\text{exp}(-\frac{t^2}{\tau_{d}^{(s)2}}),
\ee
where $P_{N-e}(t)$ is the purity decay due to intra-molecular electron-nuclear coupling (N-e), and $P_{N-s}(t)$ that due to coupling to solvent (N-s). Here, $\tau_{d}^{(e)}$ and $\tau_{d}^{(s)}$ are the corresponding decoherence times. The purity decay due to N-e interactions is usually faster than that due to N-s interactions (i.e. $\tau_{d}^{(e)} \ll \tau_{d}^{(s)}$) as a consequence of the difference in timescales of electronic and solvent motion.

In this context, it becomes clear that the origin of the apparent discrepancy is in what bath generates the decoherence. Ref.~\onlinecite{Fleming1990} considers decoherence of a vibrational state associated with a single electronic crude Born-Oppenheimer state caused by solvent. In this case, there is no appreciable entanglement between the electrons and vibrations, i.e. $P_{N-e}\approx 1$, and the purity decay due to solvent dominates i.e. $P_{N} \approx P_{N-s}$. Thus, in this scenario the vibrational decoherence would be slower than the electronic decoherence.

By contrast, consider now the case in which the vibrations entangle both with electrons and solvent, as would be the case for a molecule prepared in state \eq{eq:init0} and immersed in a solvent. In this case, the purity decay due to N-e interaction dominates as $\tau_{d}^{(e)} \ll \tau_{d}^{(s)}$ and $P_{N} \approx P_{N-e}$ for short times. Thus, in this case, the vibrational decoherence is expected to have an initial timescale for coherence loss identical to the one of electronic decoherence as the Schmidt theorem indicates, followed by a slower decay due to solvent.

\subsection{Are classical decoherence models accurate?} 
\label{subsec:accuracy}
From a practical perspective, explicit decoherence modeling requires approximate description of the system-bath dynamics. One common approximation of practical importance in molecules is to model the nuclear degrees of freedom classically\cite{Franco2008, Franco2012, Franco2013, Shim2012, ulrich2017, akimov2014, C.Tully1998, Kapral1999, Meier2004}. In this case, decoherence is captured by propagating an ensemble of quantum-classical trajectories, each one evolving unitarily, with initial conditions sampled from an appropriate classical distribution meant to mimic the initial nuclear quantum state\cite{Franco2012, Franco2013}. The corresponding ensemble average of unitary quantum-classical evolutions mimics the nonunitary evolution of the density matrix of the system. This should be contrasted with ÒtrueÓ decoherence  where a single-quantum system becomes entangled with environmental degrees of freedom and the unitary deterministic evolution of the system plus environment leads to a nonunitary evolution of the reduced density matrix of the system.

To what extent are classical bath models able to capture quantum decoherence processes in molecules?

\begin{figure}[htbp]
\centering
\includegraphics[width=0.5\textwidth]{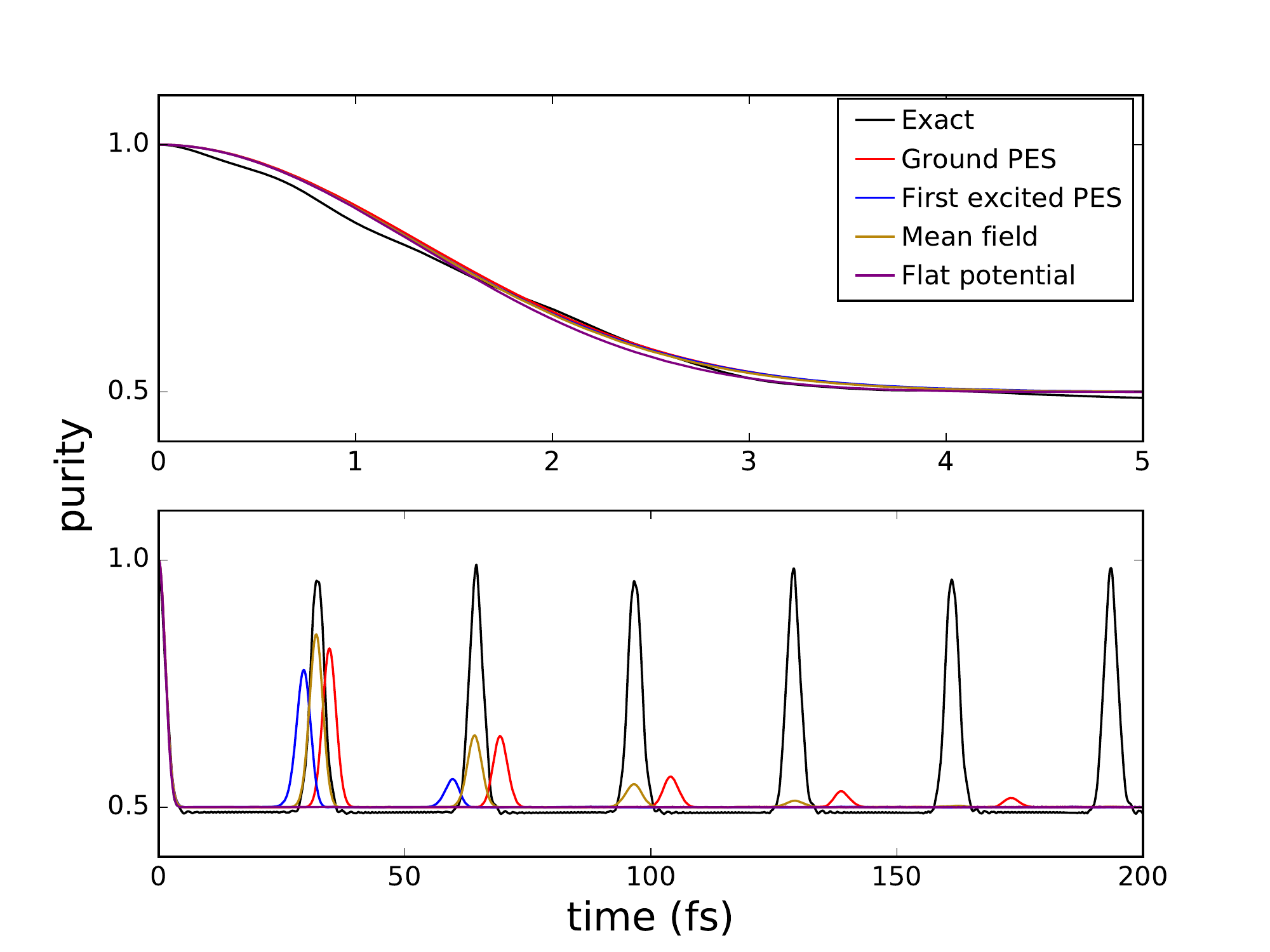}
\caption{Mixed quantum-classical description of electronic decoherence for initial state \eq{eq:init0}. The plots show the purity of an ensemble of MQC trajectories initially sampled from the ground Wigner distribution of the nuclei. The nuclei evolve in the ground (red) and first excited (blue) PES, their mean (yellow) PES and a flat potential (magenta). Note that the initial purity decay is independent of the potential on which the classical bath propagates and that it is a good approximation to the exact decoherence for short times. For longer times the MQC scheme overestimates the overall decoherence time when recurrences are present.}
\label{fig:approx_compares}
\end{figure}

To test this, we contrasted the exact decoherence dynamics of the SSH chain obtained in Sec.~\ref{subsec:emergency} as prepared in state \eq{eq:init0} with results obtained in a mixed quantum-classical approximation (MQC) where the nuclei move along a given fixed PES and the electrons instantaneously respond to the nuclear coordinates. In the MQC, we choose the initial conditions for the nuclei by sampling from the Wigner distribution of ground vibrational state of the ground PES. At time $t$, the electronic density matrix for the $i$th trajectory is:
\begin{widetext} 
\be
\label{elec_density} 
\rho_{e}^{(i)}(t) \doteq 
\begin{pmatrix} 
\frac{1}{2} & \frac{1}{2}e^{-\frac{i}{\hbar}\int_{0}^{t}{(E_{0}(u^{(i)}(t'))-E_{1}(u^{(i)}(t')))dt'}} \\
\frac{1}{2}e^{-\frac{i}{\hbar}\int_{0}^{t}{(E_{1}(u^{(i)}(t'))-E_{0}(u^{(i)}(t')))dt'}} & \frac{1}{2}
\end{pmatrix},
\ee
\end{widetext}
where $E_0(u^{(i)}(t))$ and $E_1(u^{(i)}(t))$ are the ground and first excited electronic energy at nuclear geometry $u^{(i)}(t)$. To represent the electronic density operator in matrix form as in \eq{elec_density}, it is supposed that in the region where the nuclear wavepacket is distributed the electronic states are well approximated by diabatic states (with no dependence on the nuclear coordinates) obtained by diagonalizing the Born-Oppenheimer electronic Hamiltonian at the ground-state minimum energy geometry. The quantities $E_0$ and $E_1$ are determined by diagonalizing the full SSH Hamiltonian [\eq{eq:SSH}] at fixed nuclear positions $u^{(i)}(t)$ encountered during the dynamics. The $u^{(i)}(t)$ for each trajectory is determined by solving Newton's equations of motion in a given potential $V(u(t))$. The electronic density matrix of the ensemble is taken to be the average over $N_\text{traj}$ trajectories: 
\be
\label{ave_density}
\bar{{\rho}}(t)_{e} = \frac{1}{N_{\text{traj}}} \sum_{i}{{\rho}_{e}^{(i)}(t)}.
\ee
Using the average electronic density matrix, the purity is computed as in \eq{eq:purity}. Figure~\ref{fig:approx_compares} compares the electronic decoherence dynamics generated by moving the nuclei classically along the ground PES ($V = E_0(u(t))$), first excited PES ($V = E_1(u(t))$), the mean field PES between the two ($V = (E_0(u(t))+E_1(u(t)))/2$), and a flat PES ($V = 0$) with the exact results. For short times (top panel), electronic decoherence dynamics generated by MQC along \textit{any} potential essentially coincides with the exact method. 

These simulations numerically validate, for the first time, an intriguing recent theoretical analysis\cite{Gu2017} which indicate that MQC methods correctly capture the initial purity decay when the initial state is sampled from the Wigner distribution, irrespective of the potential that is employed in the dynamics. Beyond short times, the simulations show that MQC schemes can capture some of the quantum recurrences, albeit the dynamics beyond short times depends sensitively on the potential $V$ and severely overestimates the decoherence for this model.

Thus, MQC schemes with initial Wigner sampling for decoherence are expected to be accurate in the condensed phase where the decoherence time is generally governed by the initial decay of purity. 
 
\subsection{How does electronic decoherence timescales vary with the mass of the nuclear bath?}
\label{subsec:mass_eff} 
It is challenging to intuitively predict what would be the effect of changing the mass of the nuclear bath on electronic decoherence timescales. Heavier nuclei move slower and hence  the decoherence is expected to be slower too. However, increasing the mass (decreasing the frequency) of the bath also makes the energy spectrum of the bath denser and this is expected to lead to faster decoherence. Which of these two processes is dominant? Further, what happens as the nuclear mass becomes comparable to the mass of the electron and we are beyond the regime of applicability of the Born-Oppenheimer approximation? 

\begin{figure}[htbp]
\includegraphics[width=0.5\textwidth]{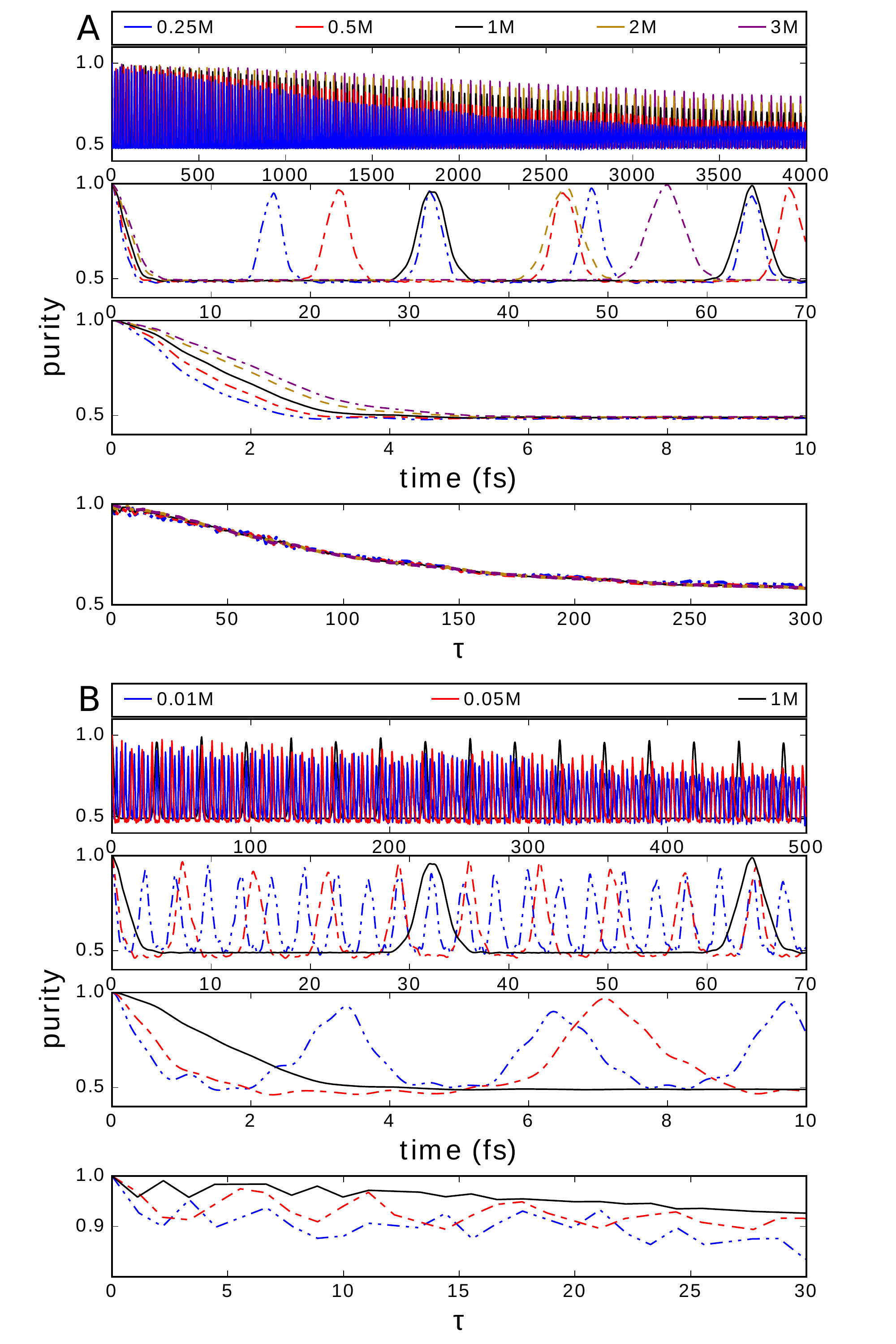}
\caption{Dependence of the decoherence dynamics on the nuclear mass in units of $M=1349.14\,\text{eV}\text{fs}^2/\text{\AA}^2$. Bottom panels plot the envelope of the purity decay (defined by the positions of the peaks in $P(t)$) versus dimensionless time $\tau=\frac{t}{2\pi}\sqrt{\frac{3K}{M}}$ for all masses.}
\label{fig:mass_eff}
\end{figure}

\begin{figure}[htbp]
\includegraphics[width=0.5\textwidth]{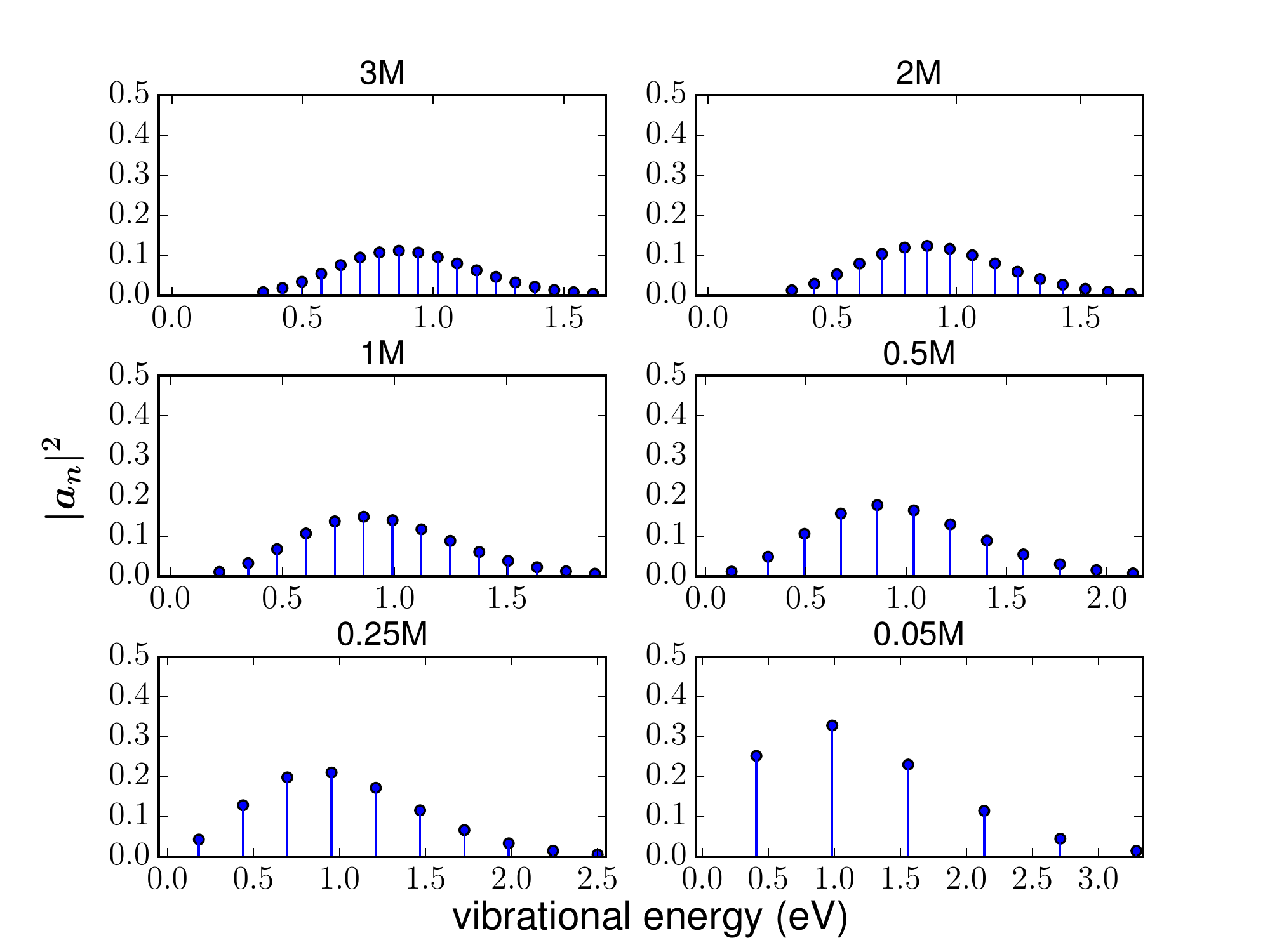}
\caption{Components $\{a_n=\langle \phi_n|\chi_1(0)\rangle\}$ of the initial nuclear state $|\chi_1(0)\rangle$ along the vibrational eigenstates $\{|\phi_n\rangle\}$ of the first excited PES. Note that the vibrational energy spectrum becomes more dense as the mass is increased.}
\label{fig:prob_distri_masses}
\end{figure}

Insights into these problems were obtained by computing the purity dynamics for SSH chains initially prepared as in \eq{eq:init0} with varying masses (0.01M, 0.05M, 0.25M, 0.5M, 1M, 2M, 3M; $M=1349.14\,\text{eV}\text{fs}^2/\text{\AA}^2$), see \fig{fig:mass_eff}. In different time windows, for 0.25M-3M, as mass decreases we observe a faster decay of initial purity decay, an earlier appearance of first recurrence peak and a faster decay to the asymptotic purity behavior. All of these observations support our first hypothesis but contradict the second hypothesis. To differentiate the effects of these two possible contributions to the purity dynamics, we eliminate the effect of the speed of nuclear bath motion on the decoherence by studying how the decoherence changes versus the number of nuclear oscillation periods by defining a dimensionless time  $\tau = \frac{\omega t}{2 \pi}$. Here, the frequency of the optical mode is used and $\omega = \sqrt{\frac{3K}{M}}$. The envelope of the purity decay versus $\tau$ for various masses is shown in the bottom panels of \fig{fig:mass_eff}. It is clear that for 0.25M-3M the different decoherence versus $\tau$ plots overlap with one another. Thus, for these masses, we conclude that the effect of the energy spectrum of the bath on the decoherence is not important, even though the vibrational energy spectrum for heavier masses are, in fact, denser (see \fig{fig:prob_distri_masses}).

As shown in \fig{fig:mass_eff}B, for even smaller masses (0.01M and 0.05M), the purity decay is more complex and deviates from that observed for the other masses as the Born-Oppenheimer approximation starts to break down. However even in this cases we observe that a lighter bath leads to faster short time decoherence.

\subsection{What is the effect of preparation on the decoherence dynamics?} 
\label{subsec:prep_eff}
In previous sections, we simulated the electronic decoherence dynamics by choosing a separable superposition state of the molecule as an initial state. This is a common strategy when defining a decoherence time\cite{Paz1993, Prezhdo1998, Batista2002, Arnold2017, Vacher2017, Franco2013, Franco2012, Franco2008}. However, experiments routinely use lasers to excite molecules and prepare the initial state. Are initial separable superpositions representative of the experimental situation? How is the decoherence dynamics affected by initial state preparation? 

\begin{figure}[htbp]
\includegraphics[width=0.5\textwidth]{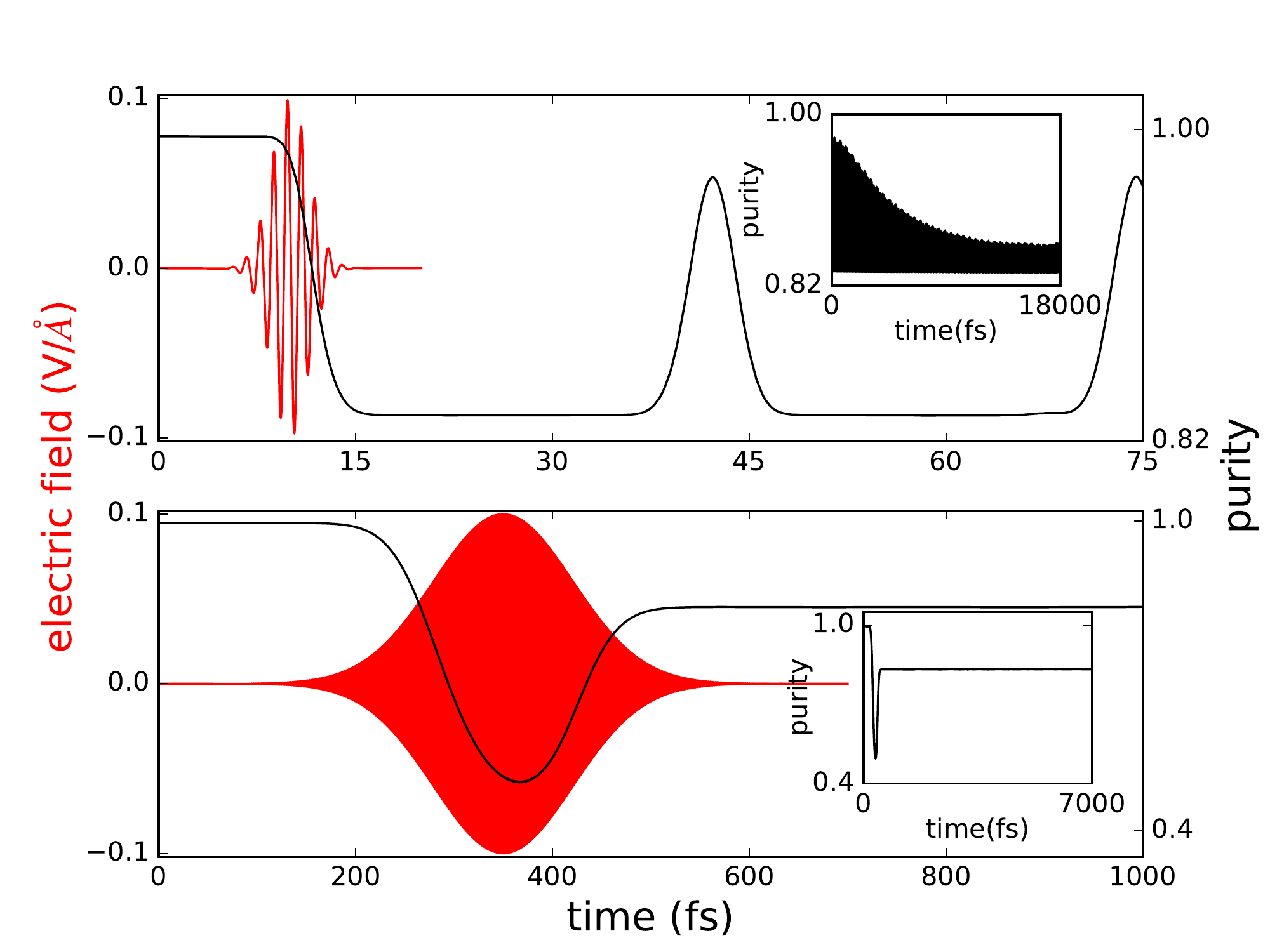}
\caption{Effect of preparation on the decoherence dynamics. The plots show purity during and after laser photoexcitation with a $2\,\text{fs}$ laser pulse (top panel) and a $100\,\text{fs}$ laser pulse (bottom panel) resonantly tuned to the $\text{HOMO}\rightarrow \text{LUMO}$ transition. Excitation via a short pulse leads to decoherence dynamics which is reminiscent to that in Fig. 1. Excitation via a long pulse is dramatically different creating an almost-stationary entangled vibronic state. Insets: long-time purity decay.}
\label{fig:laser_all}
\end{figure}

To understand this, we investigated the decoherence dynamics generated by photoexcitation with lasers of different duration. Specifically, we consider the SSH molecule in interaction with a laser in dipole approximation with radiation-matter interaction $\hat{H}_{\text{RM}} = -\hat{\vec{\mu}}\cdot \vec{\epsilon}_0 E(t)$. Here $\hat{\vec{\mu}}$ is the dipole operator of the molecule, $\vec{\epsilon}_0$ is the polarization of light and $E(t)$ is its electrical field. The laser's electric field is obtained from
\begin{align}
\label{electric field}
E(t) = - \frac{dA(t)}{dt},
\end{align}
where $A(t)$ is the vector potential defined as:
\be
\label{vector potential}
A(t) = A_{0}e^{-\frac{(t - t_{0})^2}{{t_w}^2}}\sin(\omega_0 t),  
\ee
where $A_0$ is the amplitude of the vector potential, $t_w$ is the width of the Gaussian envelope centered at $t_0$, and $\omega_0$ is the central frequency of the laser pulse. 

To study the effect of preparation on decoherence a short $t_w=2\,\text{fs}$ and a long $t_w=100\,\text{fs}$ pulse are used to resonantly photoexcite the exact ground eigenstate of the SSH Hamiltonian. The laser frequency $\omega_0$ is chosen to be at resonance with the energy difference between ground PES and first excited PES at the ground state equilibrium geometry. The frequency content of the short laser pulse ($\Delta\omega\sim 1\,\text{eV}$) can excite several vibronic transitions, while the long pulse ($\Delta\omega\sim 0.02\,\text{eV}$) can only excite a single ($0-0$) vibronic transition. 

The decoherence dynamics induced by the two pulses is shown in \fig{fig:laser_all}. Before photoexcitation, the purity is close but not exactly $1$ because the exact ground state of the molecule is entangled. The short pulse (top panel) creates a vibrational wavepacket in the excited state. In the limit in which the pulse is a delta kick and the Frank-Condon approximation is valid, this wavepacket would be identical to the one in the ground state. Since this pulse is short enough, the subsequent purity decay is reminiscent to that in \fig{fig:exact}. By contrast, the $100\,\text{fs}$ laser pulse (bottom panel) performs a state-to-state photoexcitation and no recurrences are observed. The lower purity after photoexcitation reflects the entanglement properties of the final state.

As shown, the preparation mode has a strong influence on the decoherence dynamics. Importantly, model separable initial superposition states often used to understand decoherence after photoexcitation are only relevant in the experimental situation where a short delta-like laser pulse is employed to initiate the dynamics.

\subsection{How does the electron-electron interaction affect the electronic decoherence?}
\label{subsec:e_int}

We now numerically address the largely unexplained connection between electron interactions and decoherence. According to the formal analysis in Ref.~\onlinecite{Kar2016}, electron-electron interactions can influence the decoherence rate only when the decoherence dynamics is not pure-dephasing in nature, \textit{i.e.,} when $[\hat{H}_e, \hat{H}_\text{int}] \neq 0 $. Below we numerically investigate this claim and assess the quantitative effect of changing the degree of electron repulsion on the decoherence. For this, we vary  $U$ from $0~\text{eV}$ to $0.8~\text{eV}$ and investigate its effect on the purity when the system is prepared in (i) the superpostion in \eq{eq:init0} whose dynamics is well approximated by a pure-dephasing model and (ii) a superposition of the form 
\be
\label{eq:init_ee_int}
\ket{\Omega(t=0)} = \frac{1}{\sqrt{2}}(\ket{E_0(u_{0})} + \ket{E_9(u_{0})})\otimes \ket{\chi(t=0)},
\ee
that involves states with avoided crossings and therefore is not expected to be pure-dephasing in nature. Here, $\ket{E_9(u_{0})}$ is the ninth excited electronic state determined at the equilibrium nuclear coordinates $u_{0}$ of the ground PES (see \fig{fig:pes}).

\begin{figure}[htbp]
\includegraphics[width=0.5\textwidth]{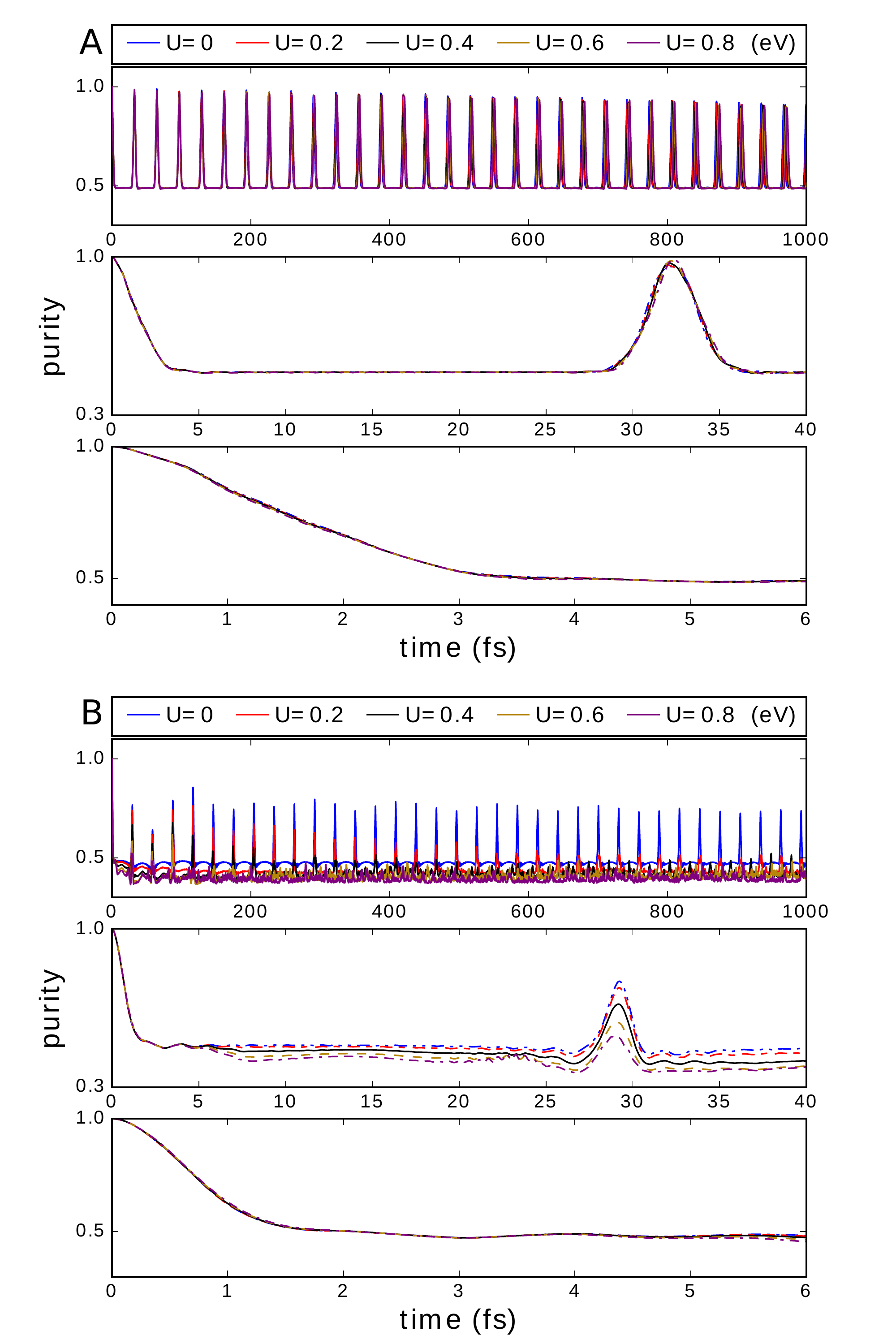}
\caption{Exact decoherence dynamics for the neutral SSH chain with 4 electrons under different electron-electron interaction strengths $U$. The chain is initially prepared in a separable superposition state between (A) the ground and first excited electronic states (\eq{eq:init0}) and (B) the ground and the ninth excited state (\eq{eq:init_ee_int}). The PESs for $U=0$ and $U=0.8~\text{eV}$ are shown in \fig{fig:pes}. Note that changing $U$ only affects the electronic decoherence in case (B) where the dynamics in not pure-dephasing.}
\label{fig:U_eff}
\end{figure}

Consider first the initial state in \eq{eq:init0}. The electronic decoherence dynamics with various electron-electron interaction strengths $U$ is shown in \fig{fig:U_eff}A. As shown, the decoherence dynamics observes minor changes when the electron-electron interaction strength $U$ is increased in this case, which is in agreement with the analysis in Ref.~\onlinecite{Kar2016}. From the perspective of the PES, the reason for this behavior is that the electron-electron interaction does not introduce significant changes to the shape of the ground and first excited PES (\fig{fig:pes}). 

By contrast, when the system is initially prepared in state  \eq{eq:init_ee_int}, changing $U$ has a strong effect on the purity dynamics. Specifically, as shown in \fig{fig:U_eff}B (bottom panel), the short-time purity dynamics is not affected by changing $U$. However, after $5~\text{fs}$ $P(t)$ changes strongly as $U$ is varied because changing $U$ changes the shape of the PESs and the magnitude of the avoided crossings involved. The fact that changing $U$ strongly influences $P(t)$ is in agreement with the formal analysis in Ref. \onlinecite{Kar2016}. The numerical observation that for the first $5~\text{fs}$ $P(t)$ is not affected by changing $U$ indicates that this segment of the dynamics is approximately pure-dephasing in nature. In fact, an analysis of the population of the diabatic states (not shown) indicates that significant population transfer to other dibatic states (4 as labeled in \fig{fig:pes}) only occurs after $5~\text{fs}$. 

Thus, changing $U$ can influence the decoherence dynamics by changing the shape of the PESs and by suppressing or enhancing avoided crossings. The numerical observations are consistent with the proposal in Ref.~\onlinecite{Kar2016} that the electron correlation and the electronic decoherence only couple for non pure-dephasing dynamics.

\section{Conclusions}
\label{sec:Conclusions}
In this work, we have presented numerical simulations of electronic decoherence in a model vibronic SSH molecule with 4 electrons and 2 vibrations using an exact method that takes into account electrons and nuclei explicitly and fully quantum mechanically. The simulations serve as a standard of electronic decoherence in molecules and address several fundamental questions about electronic decoherence.

Specially, we show that:
\begin{enumerate}
\item Coupling to one anharmonic vibrational coordinate is sufficient for electronic decoherence to emerge. While in condensed phase environments no recurrences are expected, in small molecules the decoherence dynamics from initial separable superposition states exhibits partial recurrences in the purity. 
\item While vibrational decoherence is usually considered to be slower than electronic decoherence, their early-time decoherence timescales can coincide, even in the condensed phase, when electrons and vibrations get entangled during the dynamics.
\item Mixed quantum-classical (MQC) methods with initial Wigner sampling capture the early-time decoherence correctly, irrespective of the potential employed in the propagation. This is in agreement with the theoretical developments in Ref.~\onlinecite{Gu2017} and suggests that MQC can be adequate to capture electronic decoherence in the condensed phase.
\item Decreasing the nuclear mass generally leads to faster decoherence. During the initial Gaussian purity decay, this feature was observed even for masses for which the Born-Oppenheimer approximation is expected to fail.
\item The initial preparation has a strong influence on the decoherence dynamics. Model separable initial superpositions often used to understand decoherence after photoexcitation are only relevant in experiments that employ delta-like laser pulses to initiate the dynamics.
\item Electron-electron interactions can strongly affect the electronic decoherence when the electron-nuclear dynamics is not pure-dephasing. Electronic interactions influence the decoherence by changing the shape of the PES and modulating the strength of non-adiabatic effects in the dynamics. These numerical observations agree with and give insights into the formal argument in Ref.~\onlinecite{Kar2016}.
\end{enumerate}

While the results pertain to a specific model system, the generic nature of the employed Hamiltonian permits interpreting them in a broader sense. As such, these insights can be used to interpret coherence phenomena in Chemistry and develop modeling strategies.

\begin{acknowledgements}
This material is based upon the work supported by the National Science Foundation under CHE-1553939. I.F thanks Heiko Appel, Lorenzo Stella and Angel Rubio for useful discussions.
\end{acknowledgements}
\bibliography{reference_paper1}

\end{document}